\documentclass[]{spie}  %>>> use for US letter paper
\usepackage{amsmath,amsfonts,amssymb}
\usepackage{graphicx}
\usepackage[colorlinks=true, allcolors=blue]{hyperref}

\title{Dark Energy Spectroscopic Instrument (DESI) Fiber Positioner Thermal and Wind Disturbance Test}

\author[a]{Kai Zhang }
\author[a]{Joseph H. Silber}
\author[b]{Henry D. Heetderks}
\author[a]{Daniela Leitner}
\author[c]{Michael Schubnell}
\author[a]{Michael Levi}
\author[d]{Gradey Wang}
\author[c]{Kevin Fanning}
\author[a,d]{Parker Fagrelius }
\author[a]{Carl Dobson}
\author[a]{Jessica Aguilar }

\affil[a]{Lawrence Berkeley National Laboratory, 1 Cyclotron Rd., Berkeley, CA 94720, USA}
\affil[b]{University of California at Berkeley Space Sciences Laboratory, Berkeley CA}
\affil[c]{University of Michigan, Ann Arbor, Michigan, USA}
\affil[d]{University of California at Berkeley, Berkeley CA}
\authorinfo{Further author information: (Send correspondence to K.Zhang.)\\A.A.A.: E-mail: zkdtckk@gmail.com, Telephone: 1 859 559 1811\\ }

\begin{document} 
\maketitle

\begin{abstract}
The Dark Energy Spectroscopic Instrument (DESI) is under construction to measure the expansion history of the Universe using the Baryon Acoustic Oscillation technique.  The spectra of 35 million galaxies and quasars over 14000 sq deg will be measured during the life of the experiment.  A new prime focus corrector for the KPNO Mayall telescope will deliver light to 5000 fiber optic positioners.  The fibers in turn feed ten broad-band spectrographs. To achieve this goal, it is crucial to guarantee that fiber positioners work properly under the extremes of potential operating conditions, including the full range of temperatures, high speed wind disturbance etc. Thermal testing provides valuable insight into the functionality of the fiber positioners that can be used to help mitigate poor performance at extreme temperatures and wind disturbance test provide guidance to design of ventilation system. Here, we describe the thermal and wind disturbance tests for DESI fiber positioners and how the test results helped improve the robustness of the positioners. 

%The Dark Energy Spectroscopic Instrument (DESI) is under construction to measure the expansion history of the Universe using the Baryon Acoustic Oscillation technique.  The spectra of 35 million galaxies and quasars over 14000 sq deg will be measured during the life of the experiment.  A new prime focus corrector for the KPNO Mayall telescope will deliver light to 5000 fiber optic positioners.  The fibers in turn feed ten broad-band spectrographs. We will describe the thermal and lifetime test for DESI fiber positioners and how the test results help improve the robustness of the positioners significantly. 

\end{abstract}


\section{INTRODUCTION}
\label{intro.sec}
Large spectroscopic surveys have radically changed the study of modern astrophysics. The pioneering Sloan Digital Sky Survey (SDSS, Gunn et al. 1998, York et al. 2000) delivered spectra for millions of galaxies, quasars, and Milky Way stars selected by colors and magnitudes. These precious data are released for public use, which has a huge impact on the development of astronomy all over the world. The homogeneous sample selection and large sample size enable astronomers to  do cutting-edge studies on cosmology, extra-galactic, quasar, and stellar physics. 

A key feature for large spectroscopic surveys are a multi-fiber focal plane system that retrieves light from hundreds or thousands of objects simultaneously. SDSS originally had 640 3" fibers for observation, and upgraded to 1000 2" fibers after the summer of 2009. The lights from different targets are then transmitted to the spectrographs through fibers and finally detected by CCDs. The target positions are random: for different regions on the sky, the fiber locations on the focal plane have to be adjusted. It is a time-consuming task to place fibers at the right position each night. The SDSS solution is to use cartridges to hold an aluminum plate, bend it slightly to fit the focal plane shape, and drill 640 holes at predetermined positions (Gunn \& Knapp 1993). There are 14 additional holes for guider fibers. The plates are made and drilled at the University of Washington, and shipped to Apache Point Observatory (APO) by FedEx. Each morning, technicians (pluggers) onsite plug fibers into holes, preparing for the observation at night. To determine the location of each fibers on the plate, a laser beam is transmitted through each fiber subsequently, and a camera is used to pin down the position of that fiber. 

The whole system of cartridge+plate+manual plugging of fibers for SDSS was successful. The plate designers, mill shop, FedEx, APO onsite managers, pluggers, operations managers have been working closely together to guarantee optimal use of observing time for decades. During the real operations, however, some issues arise. The plate designers need to start designing a plate 2-3 months before the planned observation time to ensure the plate will arrive on time. If some plates arrive late, significant human intervention is needed to continue overnight operations. Since plates are designed 2-3 month in advance, it is impossible to make adjustments onsite. Furthermore, training new technicians to install plates and plug fibers can be time-consuming and labor-intensive. The process of plugging in fibers inevitably bends and twists fibers, resulting in fiber, ferrule, or jacketing damage. Changes in observation plans should be done before 8:00am MDT, thus the operations manager who is responsible for overseeing the observing plan is expected to finalize the plugging request before 8:00am MDT everyday including holidays. This requirement is workable if the operations manager lives in a EST timezone, but becomes challenging if he/she lives in PST regions. Observing an increased number of objects demands a new approach altogether. 

%One solution by a group at the University of Science and Technology of China (USTC) for the development of Large Sky Area Multi-Object Fiber Spectroscopy Telescope (LAMOST, Xue et al. 2012) project is to use two steps motors in double rotation movement to place a fiber in a confined circle (Wang 2000). LAMOST has 4000 fibers, located on a circular focal plane of 1.75m in diameter. The closest distance between positioners is 25.6mm and the length of arms are both 8.25mm, resulting a patrol radius of 16.5mm. The pointing accuracy of a LAMOST fiber positioner is 40$\mu$m (0.4" on the focal plane) and the time for repositioning of fibers is about 10 minutes. 

Dark Energy Spectroscopic Instrument (DESI) pushes the large spectroscopic survey to a new regime. DESI uses the Mayall 4m telescope at the Kitt Peak National Observatory and employs 5000 dynamic fibers to capture spectra for 35 million galaxies and quasars in a five year survey projected to start in September 2019. The design requirements of DESI fiber positioner are as follows: positioner diameter $<$8mm, pointing RMS accuracy $<$5 micron, and the time for fiber rearrangement $<$2 mins. The lifetime of the positioner $>$ 200,000 moves. In this paper, we describe our teststand setup and data analysis to quantify the performance of the DESI fiber positioner. We focus on how robust the positioners perform under different circumstances such as different temperature, wind from different directions with varying wind speed etc.

\section{LBNL Fiber Positioner Features}
\label{feature.sec}
\begin{figure*}
\includegraphics[scale=0.47]{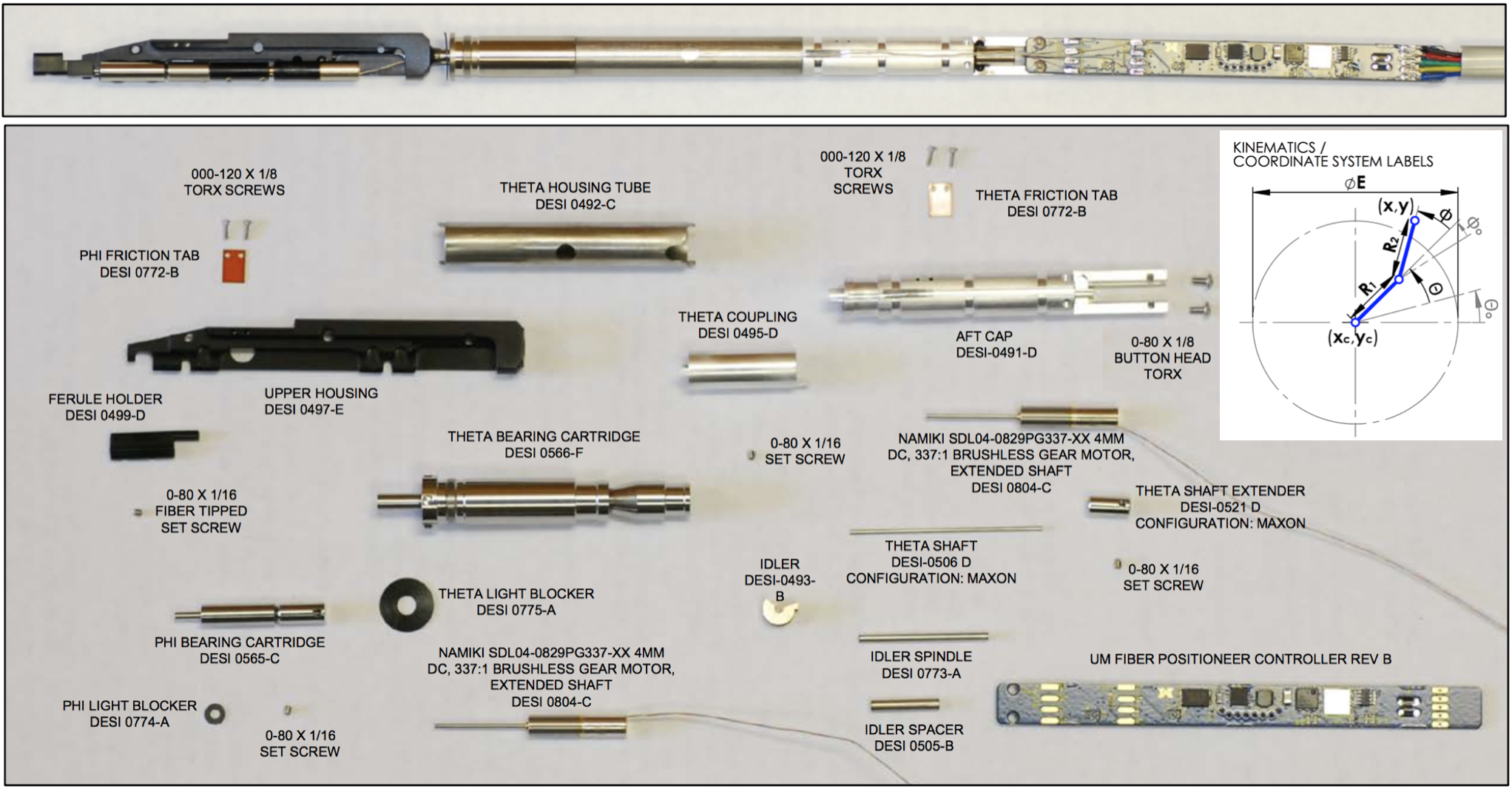}
\caption{The DESI fiber positioner sketch. The assembled positioner is shown at the top and the bottom panel shows all the parts. The names of part and their corresponding DESI document number are given. The embedded panel shows the kinematics/coordinate system. The phi angle determines the opening angle of the two arms, and is therefore analogous to the elbow joint. The theta motor is analogous to the shoulder joint because the theta angle determines the rotation of the positioner.    }
\label{fp.fig}
\end{figure*}

A picture of DESI fiber positioner is given in Figure~\ref{fp.fig}. The 22 different components are marked on the figure. The principle of this design is similar to the human arm: the phi angle determines the opening angle of the two arms, and is therefore analogous to the elbow joint. The theta motor is analogous to the shoulder joint because the theta angle determines the rotation of the positioner. The length of the two arms are 3mm. Two 4mm brushless motors made by Namiki Precision Jewel Co., Ltd  rotate the two arms respectively. Hard stops are implemented so that phi and theta can only rotate by about 180 and 390 degrees. This ensures the fiber will never be twisted to the point of failure. The closest distance between positioners is 10.4mm. A positioner control board with 5 wires provide power to the motors, transfer and process command through a CAN\footnote{Controller Area Network} BUS. Assembling a positioner requires 20 parts and 10 fasteners (Aguilar et al. 2018). The design of this fiber positioner and its performance is given in Silber et al. (2018) and Schubnell et al. (2018). 

\section{Teststand Setup}
\label{variation.sec}

\subsection{Positioner Accuracy Test ( "XY test")}

We design "XY tests" to quantify the pointing accuracy of a positioner. Each positioner is extensively tested to quantify its positioning accuracy performance. The test is a mimic of nighttime DESI operations, as well as performing sufficient moves to eliminate infant mortality of the positioners before they are installed in the telescope. We generate a series of targets evenly distributed in the patrol area of positioner, and move the positioner to the target positions. The current theta and phi angles are always tracked internally. The moving code record how much angles the arms have rotated, and get the current location based on the sum of previous moves from starting point. After taking a picture of current location of fiber tip and compare it to the target, a correction is calculated, and the positioner makes the correction move. Note the current theta and phi are not UPDATED based on the correction. The measured correction is added to the existing theta and phi values and make a correction move. The advantage of this open-loop controlling and tracking method is the pointing error only depends on the starting point of a move sequence. The disadvantage of this method is that the internal tracking could deviate from the real location if there is some issues like: loss of steps of motor, accidental collision of positioners, power outage etc.  

At the beginning of each test, a calibration procedure is performed to determine the center of the fiber positioner, lengths of both arms, physical ranges of theta and phi etc.  An XY test includes the following steps: \\
1) Command the positioner to move to a target position at speed of 178 deg/s. \\
2) Take an image of the illuminated fibers with the Fiber View Camera (FVC) to get the actual position of the fiber.  Correction move angles are calculated based on this. \\
3) Make a correction move according to the derived corrections at speed of 2.6667 deg/s. \\
4) Repeat 2) and 3) N times. N=4 for our test shown here.  \\
DESI pointing error requirement is pointing error RMS$<$6$\mu$m. %We give detailed final positioner's performance in Section~\ref{performance.sec} 
We typically achieve 5$\mu$m accuracy with 1-2 correction moves for most of our final production positioners. The mean blind move positioning error was measured to be less than 30$\mu$m and the mean RMS for the correction move was measured to be 1.6$\mu$m (Schubnell et al. 2018).

\begin{figure*}
\includegraphics[scale=0.185]{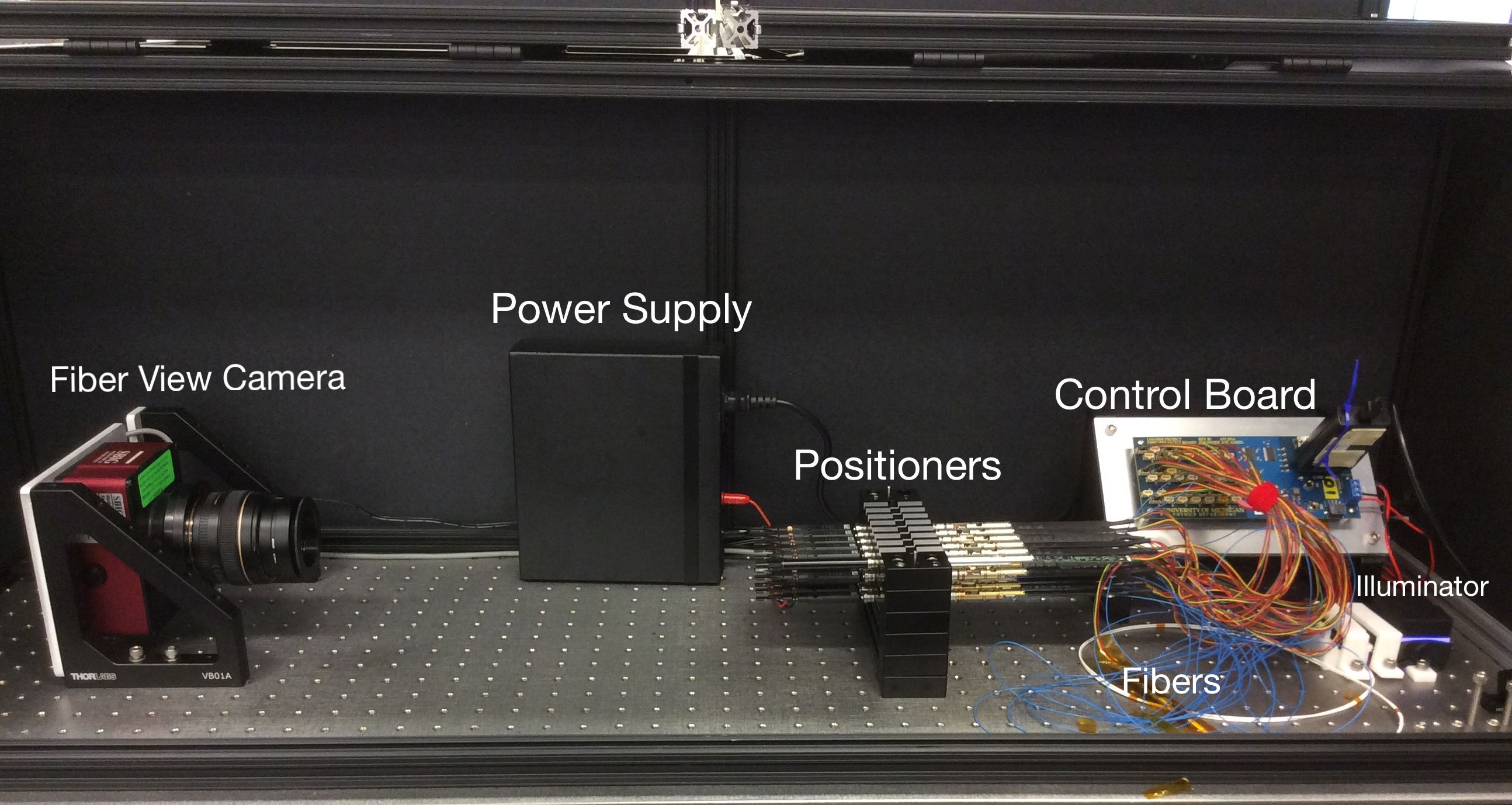}
\caption{The teststand for XYtest at LBNL.  }
\label{LBNL1.fig}
\end{figure*}

\begin{figure*}
\includegraphics[scale=0.185]{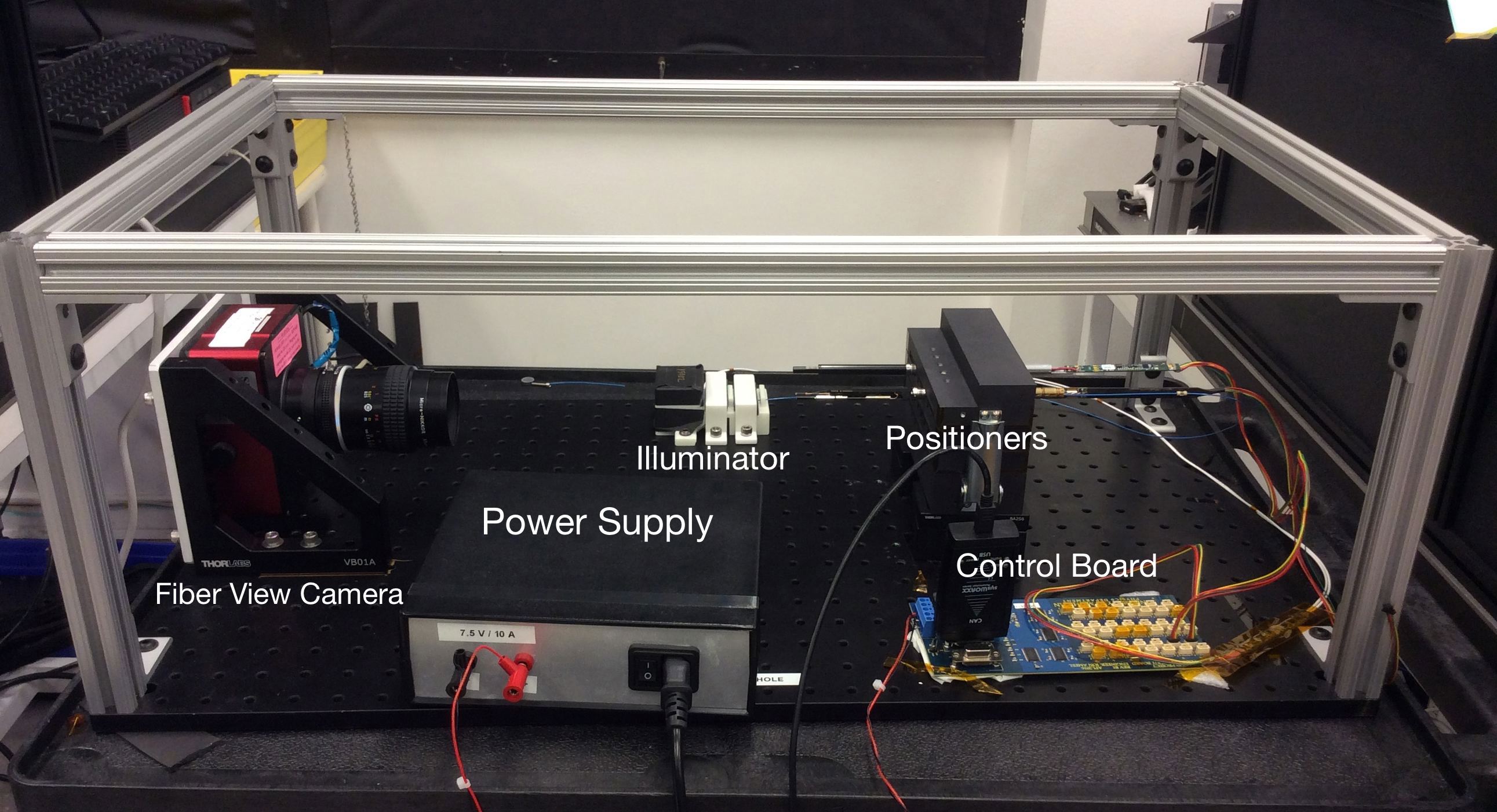}
\caption{The teststand for thermal test at LBNL.   }
\label{LBNL0.fig}
\end{figure*}

The hardware setup of the teststand to implement the XYtest is shown in Figure~\ref{LBNL1.fig}. We use a SBIG 8300M camera with a lens ( For example, Nikon Micro-NIKKOR 55mm f/2.8 Lens) mounted as the Fiber View Camera (FVC). For the final DESI system, we adopt a Kodak KAF50100 CCD with 6132$\times$8176 pixels 6$\mu$m by 6$\mu$m each, operated at room temperature. A Canon EF 600 mm f/4 lens, with a Canon lens mount the lens focusing and f stop can be remotely control. A Proline PL50100 controller available from Finger Lake Instruments is employed as CCD control and readout electronics. A clamp rig mount with similar thickness (6.8cm) as the petal is used to hold the fiber positioners. Both the camera and the stand are fixed to an optical bench, and this distance differs across different teststands. %All the instruments will eventually be controlled by DESI Online System (DOS), which connects and communicates with individual subsystems. 
A small single-board computer (Beaglebone Black) sends commands to positioners, fiducials, and illuminator through a CAN bus. A PC runs the python test code, which point positioners to targets in close loop, take an image to pin down where the positioners really located, and send correction move commands to positioners.  We run 3-4 corrections moves before moving to the next set of target. 

An XY test grid could contain arbitrary number of targets. We use 192 targets for reference. The following statistics are used to quantify the performance of positioners in a given run:

\begin{description}
\item [(1)] Blind Move Maximum Offset (Blind Max): describes the largest offset in a test when we move the positioner to a new target in open loop. This number defines the collision radius for each positioner. \\
\item [(2)] Correction Move Maximum Offset with 5$\mu$m threshold (Corr Max): describes the largest final offset after doing multiple correction moves. We stop correction after reaching the threshold of 5$\mu$m. 
\item [(3)] Correction Move RMS (Corr RMS): The root mean square of the final offset. 
\item [(4)] Correction Move Offset with 5$\mu$m threshold Best 95\% (Corr Max Best95): describes the largest final offset after doing multiple correction moves for the best 95\% grid points. 
\item [(5)] Correction Move RMS Best 95\%(Corr RMS): The root mean square of the final offset for the best 95\% grid points. 
\end{description}
These parameters are used for the grading of the fiber positioners. The details of the grading criteria are given below: 
\begin{description}
\item Grade A: Blind Max $<$100$\mu$m, Corr Max$<$15$\mu$m, Corr RMS$<$5$\mu$m. 
\item Grade B: Blind Max $<$250$\mu$m, Corr Max$<$25$\mu$m, Corr Max 95$<$15$\mu$m, Corr RMS$<$10$\mu$m, Corr RMS 95$<$5$\mu$m. 
\item Grade C: Blind Max $<$250$\mu$m, Corr Max$<$50$\mu$m, Corr Max 95$<$25$\mu$m, Corr RMS$<$20$\mu$m, Corr RMS 95$<$10$\mu$m. 
\item Grade D: Blind Max $<$500$\mu$m, Corr Max$<$50$\mu$m, Corr Max 95$<$25$\mu$m, Corr RMS$<$20$\mu$m, Corr RMS 95$<$10$\mu$m. 
\item Grade F: Positioner does not meet any of the above. 
\end{description}

For each positioner, a burn-in test composed of several individual XY tests is performed right after it is fully assembled. The burn-in test run 24 points test at 40\%, 50\%, 60\%, 70\%, 80\%, 90\%, 100\% current, and a 192 point XYtest at 100\% current. The final grading is based on all test done at current $>=$ 70\%. The classification of positioners into Grades A through F were done at the University of Michigan (Schubnell et al 2018).
The yield of Grade A and B positioner, which are installed into production petals, are 96.1\% and 0.8\% after Sep 2017. The details of DESI fiber positioner accuracy, speed of convergence, lifetime and production yield are given in Schubnell et al. (2018). 

\subsection{Thermal Test }
\label{s2.sec}
\begin{figure*}
\includegraphics[scale=0.4]{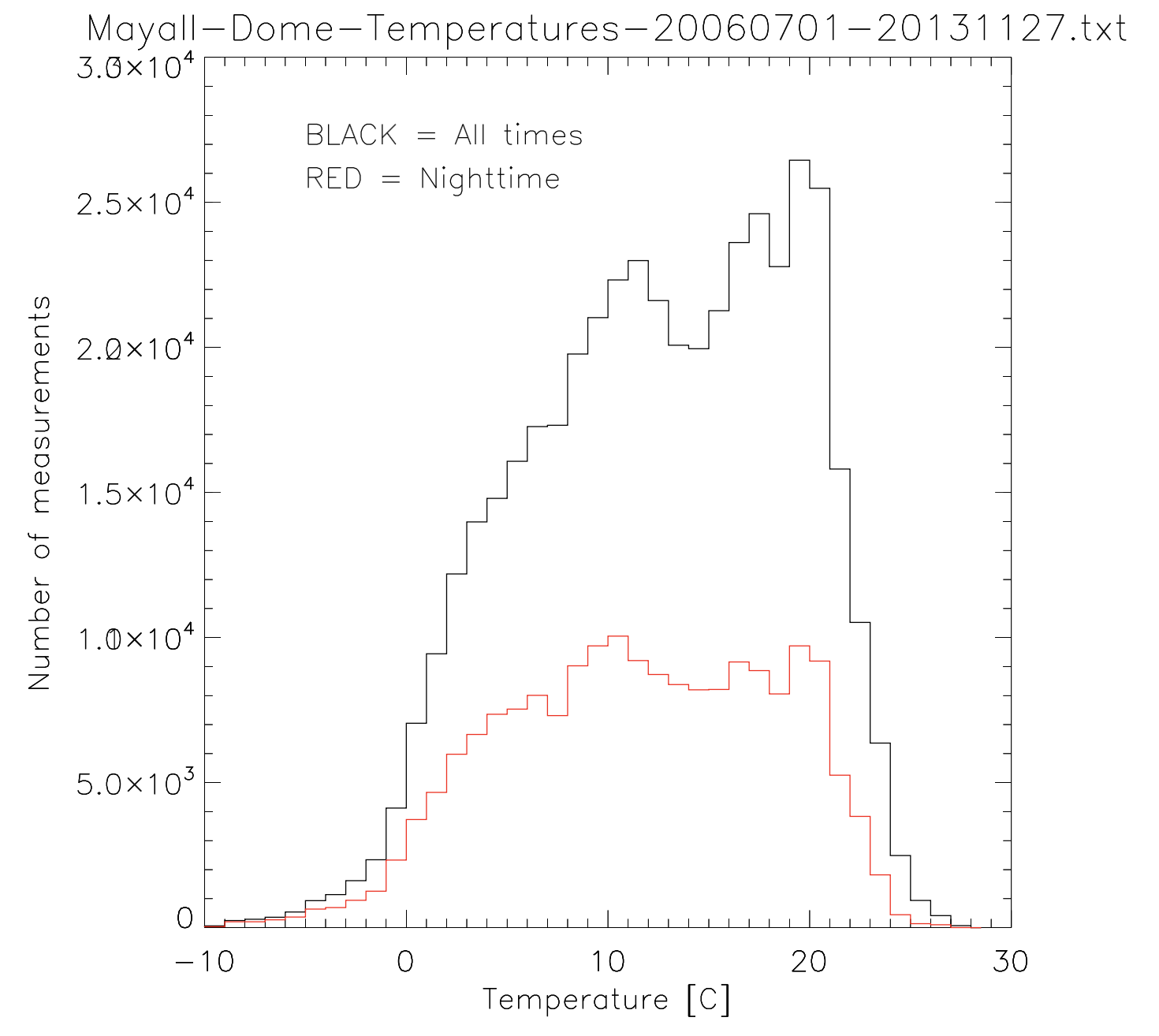}
\caption{Mayall dome temperature histogram. The data is taken between July 1st 2006 and November 27th 2013.  }
\label{temperature.fig}
\end{figure*}

\begin{figure*}
\includegraphics[scale=0.4]{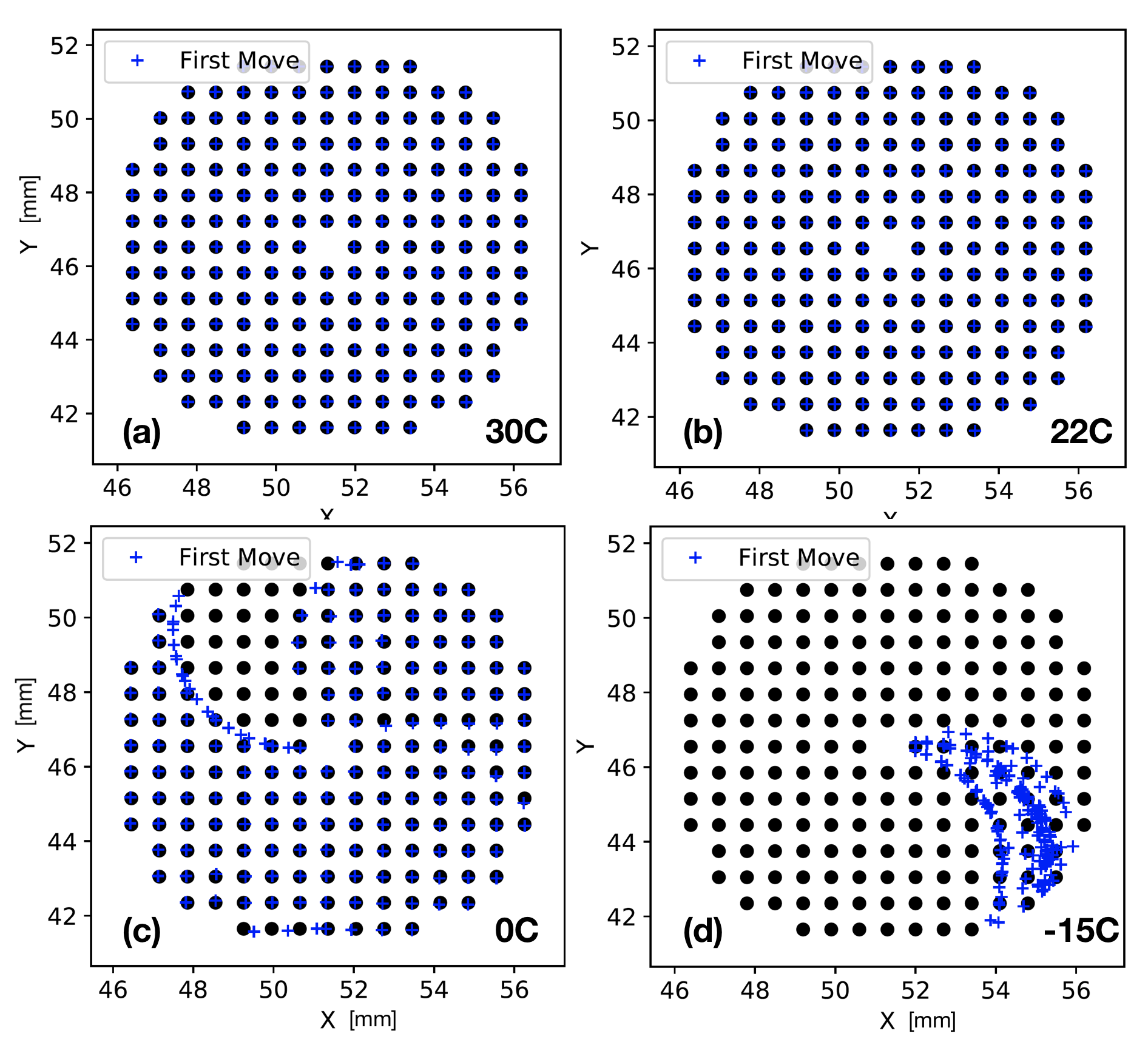}
\caption{The XY test blind move performance plots for positioner M01012. The black dots are the target positions, and the blue crosses are the actual fiber positions after the first blind move. For the upper two panels, the blue crosses are overlapped with the black dots. Panel (a), (b), (c), (d) are for T=30$^{\circ}$C, 22$^{\circ}$C, 0$^{\circ}$C, -15$^{\circ}$C respectively. The theta motor is stuck at T=-15$^{\circ}$C, while phi motor can still rotate. In this case, the actual positions of fiber would locate in an circle with a radius of arm length (3mm). This is exactly what we see in Panel (d). When temperature rises, the motor/gear box work again.  }
\label{thermal_xy_1012.fig}
\end{figure*}

The night temperature at Kitt Peak varies between -10$^{\circ}$C and 30$^{\circ}$C as shown in Figure~\ref{temperature.fig}. Our fiber positioners should be able to work smoothly at any temperature within this range to guarantee good fiber positioning. Furthermore, the shift of fiber tip relative to the CCD should be quantified and understood to ensure good positioning at all temperature. For this purpose, we constructed a teststand that can be put into a thermal chamber at LBNL. The image of this teststand is shown in Figure~\ref{LBNL0.fig}. The dimensions of this teststand measure 76cm$\times$45.5cm$\times$28cm. The distance between the lens and tip of fiber positioners is about 20cm. Twelve positioners can be tested in one run, limited by the field of view of the camera and lens. We run XY accuracy tests at -15$^{\circ}$C, 0$^{\circ}$C, 22$^{\circ}$C, and 30$^{\circ}$C. For the run on 2017-06-10, nine out of twelve fiber positioners tested show very similar performance at different temperatures. They are classified as Grade A at all tested temperature. However, one positioner (M00437) worked well at first, had its phi motor stuck after 40,000 moves. Another two of the twelve positioners (M01012 and M00872) were stuck in both theta and phi directions when cooled to -15$^{\circ}$C judging from the XY test data. The upper-right panel in Figure~\ref{thermal_xy_1012.fig} shows the XY accuracy offset plot at room temperature of 22$^{\circ}$C for positioner M001012, and the bottom-right panel is the accuracy plot at -15$^{\circ}$C. The black dots are the target positions, and the blue crosses are the actual fiber positions after the first blind move. Panel (a), (b), (c), (d) are for T=30$^{\circ}$C, 22$^{\circ}$C, 0$^{\circ}$C, -15$^{\circ}$C respectively. The theta motor can reach larger area than at T=-15$^{\circ}$C, but there are still some regions it can not reach. This can be seen clearly on the calibration data taken when we ask the theta and phi motor to rotate by certain angles.  When we increase the temperature to 22$^{\circ}$C or 30$^{\circ}$C, both theta and phi motors move smoothly. M00872 shows similar behavior to M01012. 

\begin{figure*}
\includegraphics[scale=0.7]{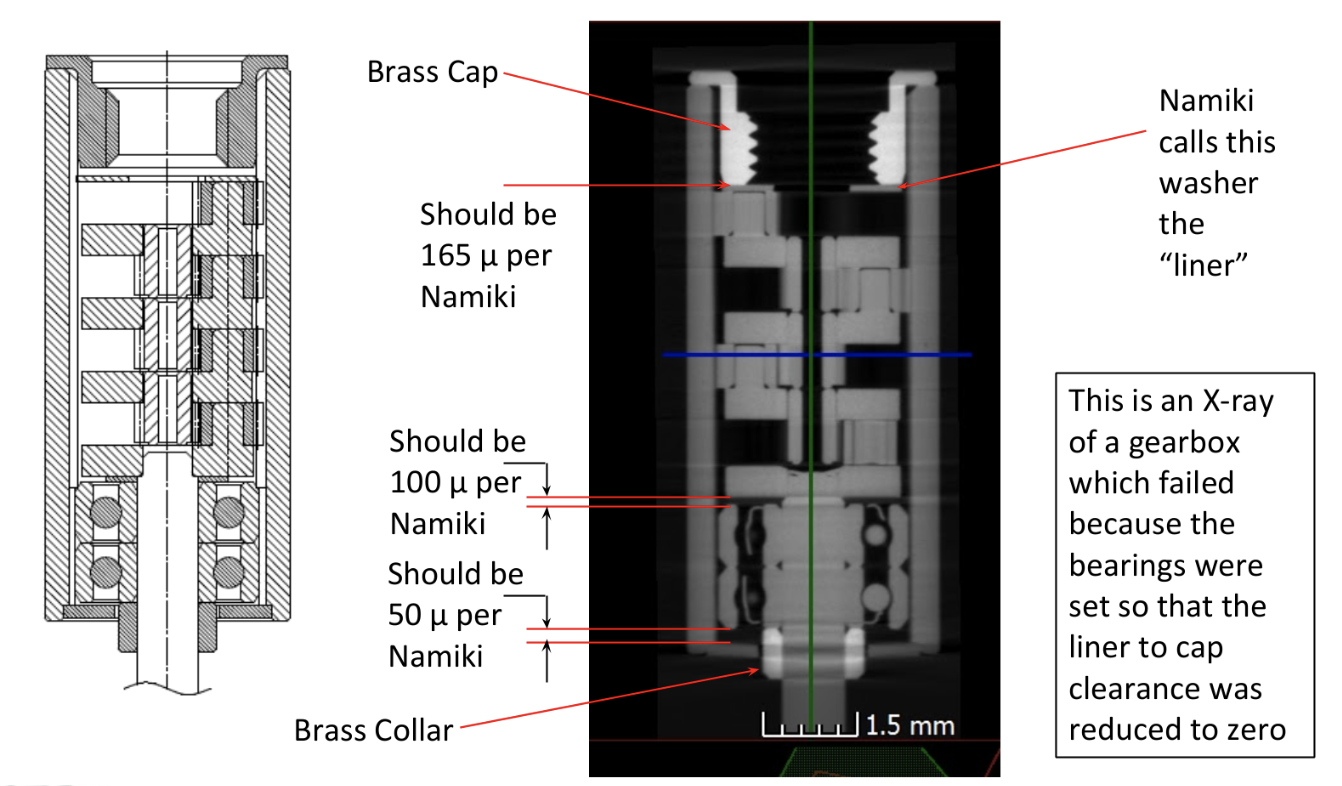}
\caption{Left: Design of the gear box for reference. The gear box and the motor are one piece, DESI 0804-C in Figure~1. Right: The X-ray scan of the cross-section of the gear box which is stuck during the thermal test.  There should be a 165$\mu$m gap between the brass cap and the washer to enable the gear box to work properly. But in this case, this gap is reduced to almost 0, because the shaft (at the bottom) compresses the gear stack.  }
\label{xray.fig}
\end{figure*}

To diagnose the cause of the stuck issue,  the three positioners that stuck at low or even room temperature were sent to X-ray scans at Los Alamos National Lab. 
The cross section of the gear box of positioner M00437 is shown in Figure~\ref{xray.fig} Surprisingly, the slack of the gearbox was reduced to nearly 0$\mu$m, while the designed value is 165$\mu$m. Without enough slack, the gears can not rotate freely. Among all the gearboxes, about 70\% of them have a motor shaft that easily wiggles along the direction of its axis. The remaining 30\% have tight shaft. 

\begin{figure*}
\includegraphics[scale=0.6]{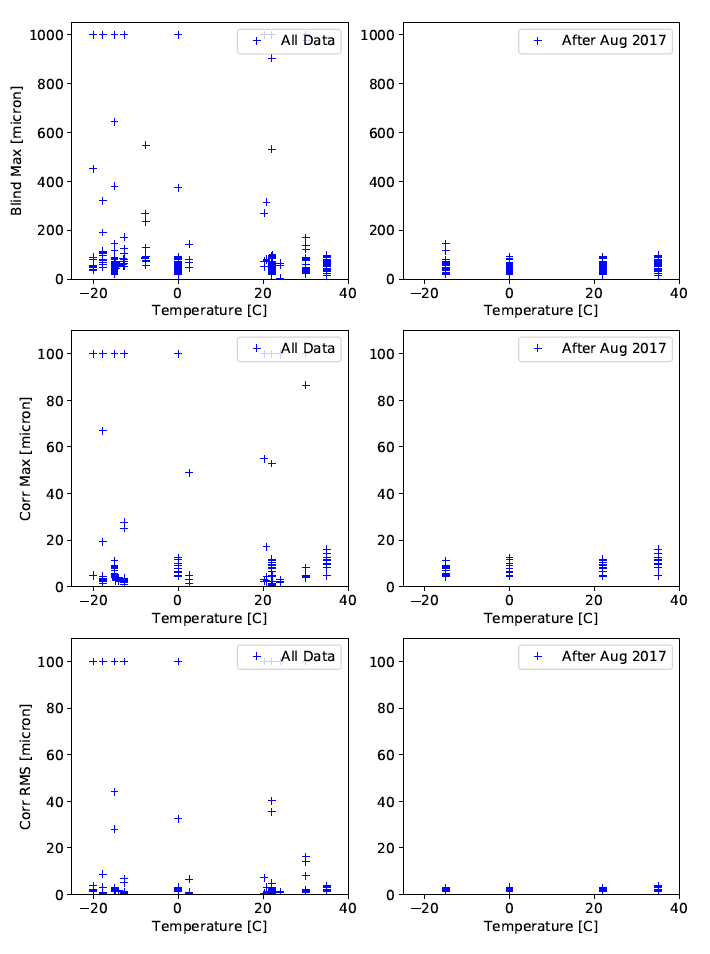}
\caption{The statistics of critical positioner performance parameters as a function of environmental temperature. }
\label{par_temp.fig}
\end{figure*}

Based on these data, we implement the following fix in assembly procedure: all shafts are intentionally biased outward when assembled. See Aguilar et al. (2018) for more details of assembly procedures. We assemble 12 positioners with tight shaft both in theta and phi, 12 positioners with wiggling shaft biased inward (thus compress the gear stack as much as the tolerance of the parts permitted) and 12 positioners with wiggling shaft biased outward (maximum clearance available) to test our hypothesis that the gearbox clearance is the reason for this failure mode. Three out of twelve positioners with shaft biased inward failed right after assembly, and another 3 failed after a burn-in test. By comparison, all the positioners with outward biased shaft and tight shaft work well under all temperatures. See Table~\ref{out_and_fix.tab} for details. We gather all the thermal test data we obtain and plot Blind Max, Corr Max, Corr RMS as a function of temperature in Figure~\ref{par_temp.fig}. The left column is for all thermal test data for the 92 positioners tested, and the right column is exclusively using the data for 32 positioners tested after we controlled the clearance in gearbox. Before fixing this issue, a significant fraction of positioners did not work well at low temperature or even room temperature. The Blind Max, Corr Max, Corr RMS are abnormally high for these positioners. However, after August 2017, Blind Max, Corr Max and Corr RMS rarely exceed 100$\mu$m, 10$\mu$m and 5$\mu$m respectively now.  
Please refer to Schubnell et al. (2018) for an evaluation of the overall performance of positioners and lifetime test result and to Aguilar et al. (2018) for the manufacturing processes.. 

%All these positioners are tested to 20,000 moves, without showing significant degradation. The grading of the 20 positioners as a function of moves they have made is given in Figure~\ref{grade_lifetime.fig}.  Overall, all of them remain Grade A throughout their lifetime. For some tests for some positioners, the performance degrade, and they are classified as Grade B or even Grade F. Two quantities: Blind Max and Corr Max could drive the grading lower with just one single bad point in a test, and there are many possibilities for producing single point deviation, like shaking of the teststand caused by the vibration of refsrigiator, temporary high friction at a specific location; temporary current change etc. This does not hide the truth that the positioners retain good performance throughout their lifetime. 

%\begin{figure*}
%\includegraphics[scale=0.7]{par_lifetime.png}
%\caption{The statistics of critical parameters as a function of number of moves the positioners have made. The xytest data for all 20 positioners in Table~\ref{out_and_fix.tab} are plotted. Generally, the performance of all positioners is very stable, no systematic in degradation in any parameter is seen.  }
%\label{par_lifetime.fig}
%\end{figure*}

\subsection{Impact of Wind}
\begin{figure*}
\includegraphics[scale=0.55]{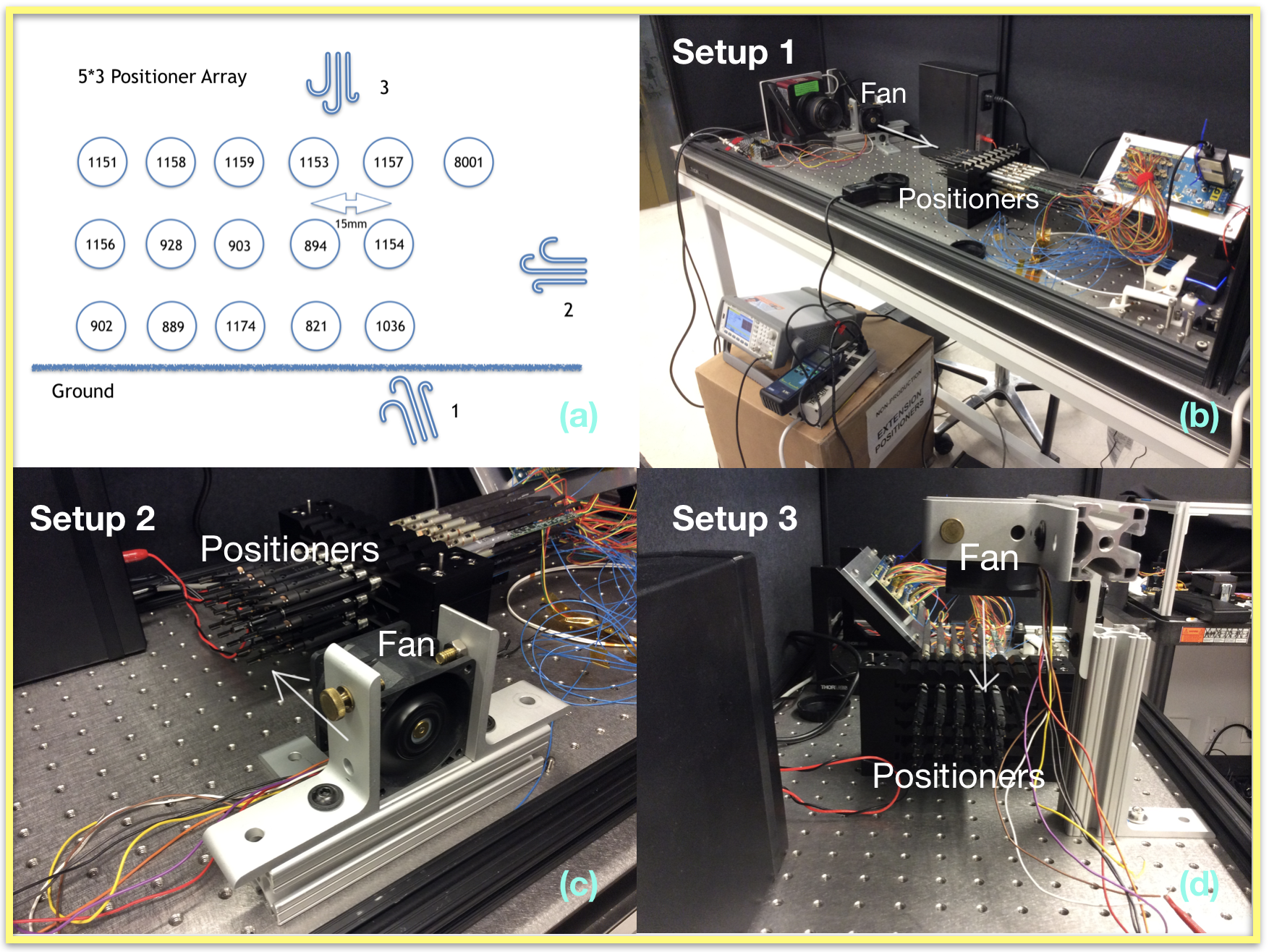}
\caption{The setup of the teststand to test the impact of wind on positioner performance. The teststand setup is identical to Figure~\ref{LBNL1.fig}, and a fan is used to blow wind to positioners in different directions. Panel (a): a sketch of the configuration of 5*3 positioners array. Each circle represents a fiber positioner, and the number is their IDs. The upper-right one: 8001 is a fiducial for registering the fiber positioner during moves. Panel (b): Setup 1 with face-on wind blowing toward the positioners. Panel (c): Setup 2 with side wind blowing parallel to the ground. Panel (d): Setup 3 with side wind blowing toward the ground.  }
\label{test_wind_setup.fig}
\end{figure*}

\begin{figure*}
\includegraphics[scale=0.7]{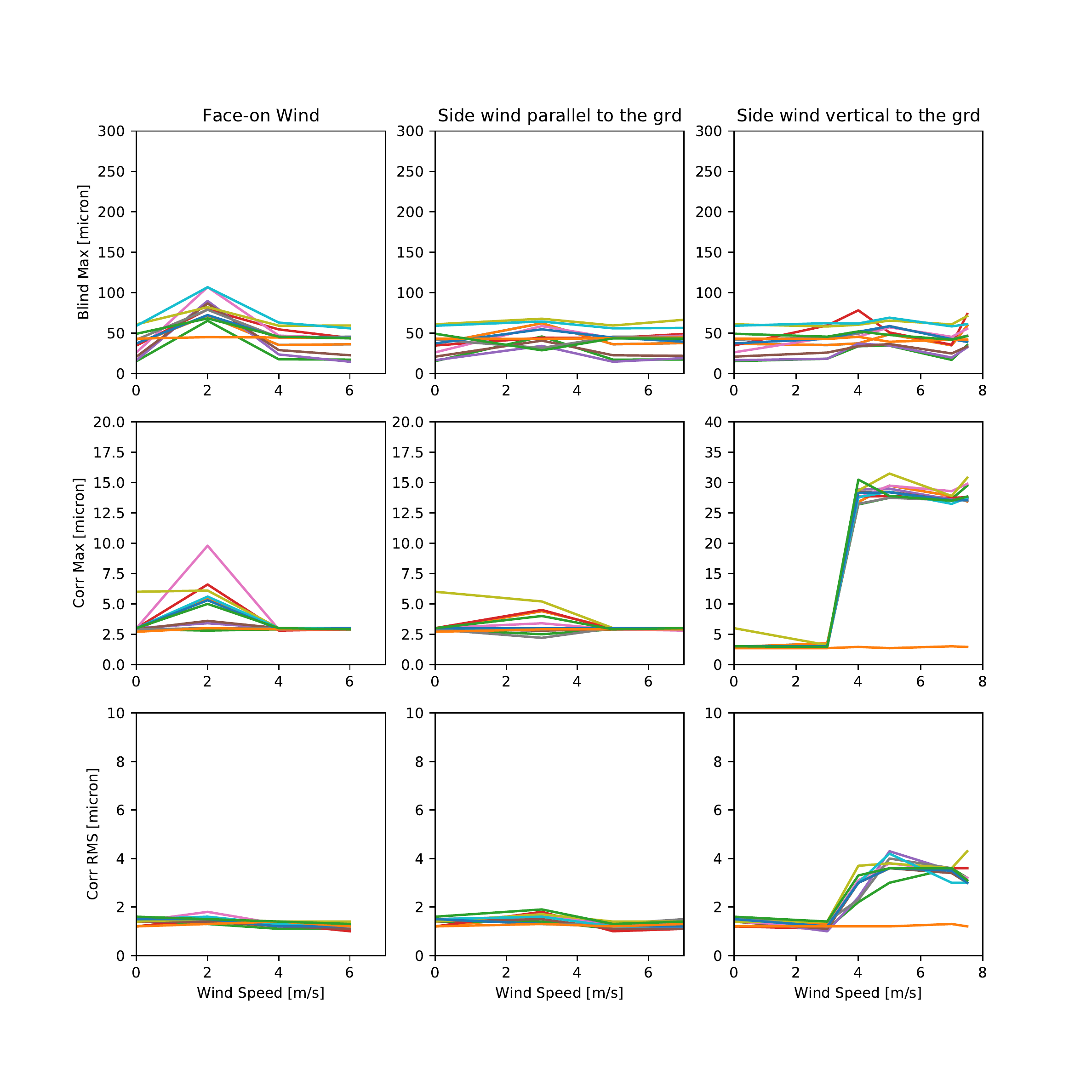}
\caption{Blind Max, Corr Max, Corr RMS as a function of wind speed for the three setups. Different colors represent different positioners. For setup 1 and 2, the performance are as good as no wind performance. For setup 3, the performance worsen starting from wind speed equals 4m/s. Between 4m/s and 7.5m/s the three parameters stay constant.  }
\label{pos_wind.fig}
\end{figure*}

To eliminate the possibility of fluid leaks within the focal plane enclosure, the protect decided to use air flow to cool the positioners. To validate the design concept  we tested how the wind impact the performance of positioners. For this purpose, we run XYtest with wind blowing in different directions at different speeds. Fan is mounted to generate wind blowing in the following directions: 1) face-on wind, 2) side wind blowing parallel to the ground, and 3) side wind blowing vertical to the ground. The three experimental test setups are shown in Figure~\ref{test_wind_setup.fig}. We run 192 points XYtest at wind speeds ranging 0 and 7m/s. Fifteen positioners were tested. The pitch between positioners was 15mm. The blind max, corr max, corr rms as a function of wind speed for these three setups are shown in Figure~\ref{pos_wind.fig}. We show the data for 12 Grade A positioners only thus reject positioners M00928, M00894, and M01036. We can see that for setup 2, the performance is as good as no wind performance at all wind speed. For setup 1, there is a jump in blind max and corr max at 2m/s. At even higher wind speed, however, blind max and corr max drop back to no wind values. It is highly possible some resonance emerges at wind speed $\sim2m/s$.  For setup 3, the performance worsen starting from wind speed equals 4m/s, and stablizes at a constant blind max, corr max, and corr rms all the way to 7.5m/s. The orange line which shows a constant behavior at all wind speed is M01174. It is located at the bottom center of the positioner array, thus probably experiences most benign wind disturbance.  All the behaviors listed are repeatable, and independent of wind speed history. Probably this is because the wind generates strong enough turbulence at 4m/s near the ground, and the turbulence vibrate the moving parts of the positioners like fibers, light blockers, etc. This test provide valuable guidance for ventilation of the positioners. First, we need to avoid the setup and wind speed that shows resonance. Second, we need to pay close attention to the positioners that located near solid walls. This is where turbulences most likely to appear and impact the performance of positioners.  

\section{Summary}
DESI positioners are now in full production. We use XYtest to quantify their performance using several statistics. For the three critical parameters: blind max, corr max, corr rms, we show they are constant from -15$^{\circ}$C to 35$^{\circ}$C. At 2m/s when wind is blowing toward the positioner, some resonance appear. When a side wind is blowing against solid surface, significant degradation in pointing accuracy is observed when wind speed exceed 4m/s. The test at extreme temperature and wind situations illustrate the robustness of DESI positioners, and give guidance to the design of ventilation system. 

%=============== Table ====================
\begin{table*}[h]
%\centering
\footnotesize\setlength{\tabcolsep}{1.8pt}
\topmargin 0.0cm
\evensidemargin = 0mm
\oddsidemargin = 0mm
\scriptsize %\small %\tiny %
\caption{Performance of Fiber Positioners with outward biased shafts and fixed shafts}
\label{out_and_fix.tab}
\medskip
\vfill
\begin{tabular}{c c |  c c c c  |c c c c  |c c c c   |c c c c }
\hline \hline
POSID&Biased	&Blind Max &  &&  &Corr Max & && &       Corr RMS &  & & \\
           &            &   (30$^{\circ}$C) &  (22$^{\circ}$C)  & (0$^{\circ}$C) & (-15$^{\circ}$C)    & ( 30$^{\circ}$C) & (22$^{\circ}$C) & (0$^{\circ}$C) & (-15$^{\circ}$C) & (30$^{\circ}$C) & (22$^{\circ}$C) & (0$^{\circ}$C) & (-15$^{\circ}$C) &\\
1151	&out&	24	&27.5&	28.1&	33.1&		8.1&	5	&6	&5.8		&	   1.9	&1.5&2.4&2.1 \\
1152	&out&	73.3&44.1 &	43.3	& 40.8	&	5&	4.9&	4.9&	5	&	   1.8&1.6&1.8&	2 \\
1153	&out & 93	&91.5	&91.6	&79.9	&	4.9	&4.8 &	7.9&	5 &	   2.1&2.1&2.3&	2.2\\
1154	&out&66.3	 &70.9	&68.6	&68&		9.9&	6.4	&12.4	&4.8 &	  2.9&2.1&	2.5&1.8\\	
1155	&out & 35.4	&25.5	&32.2&	26.9&		16.1&	9.3&	11.6&8.5	&   3.6	&2.5&3&2.4\\
1156	&out& 38.1	&42.2	&39.6&	44.6	&	4.7&	5	&5	&4.9	&	  1.6	&1.6&1.7&1.8 \\
1157	&out& 74.2	&49.1&	49.6&	48.8	&	5	&4.7&	4.9&	4.7 &	 1.9&	1.3 &1.8& 1.7\\	
1159	&out&	32.8&	24.1&	34.3&	39.1&		4.9&	5	&5	&5 &	  1.8&2&1.7& 1.7 \\
1171	&fixed&	39.7&	18.8&	20&	       21.3&		12.7&	9.6&	5&	5	&		3.2&	2.7&	2.1&	2\\							
1172	&fixed&	58.7&	62.3&	63.5&	58.4&		4.9&	4.98&	5&	8.9	&		2.2&	2.2&	2.3&	2.7\\								
1173	&fixed&	58.2&	57.9&	61.2&	71.3&		11.2&	11.8&	9.3&	9.3	&		3.1&	2.8&	2.3&	2.8\\								
1174	&fixed&	64.1&	61.8&	68.4&	63.6&		14.5&	11.2&	10&	11.2		&		3.9&	2.8&	3.2&	3\\						
1175	&fixed&	50&	         31.4&	36.8&	36.5&		5 &	4.9&	5&	5 &		2.2&	2.2&	2.1&	2.1\\								
1176	&fixed&	80.6&	56.3&	47.4&	53.9&		5 &	4.6&	5&	4.7  &		2.1&	1.9&	2.2&	2.3\\								
1177	&fixed&	41.5&	47.9&	45.8&	45.2&		9.7&	7.7&	9&	7 &		3.2&	3&	3&	2.6\\									
1178	&fixed&	76.2&	54.7&	53.1&	55.3&		8.2&	8.5&	4.9&		4.9	&		2.2&	2&	1.9&	1.9\\							
1180&fixed&	54.5&	56.2&	56.9&	61.2&		5 &	4.9&	4.9&	4.3	&		1.8&	1.7&	1.7&	1.6 \\							
1181&fixed&	35.2&	33.2&	32.4&	35.1&		4.8&	   5&	4.6&	4.8		&		1.6&	1.6&	1.4&	1.7\\							
1182	&fixed&    70.6&	48.8&	49.3&	61.5&		4.7&	   5&	4.9&	4.7		&		1.8&	1.9&	1.9&	2\\								
1183	&fixed&    97.2&	86.9&	84.2&	80.1&		11.4& 9.6&	6.5&	7 &		2.9&	2.4&	2.1&	2.2 \\
\hline

 \hline
\end{tabular}
\medskip
\vfill
\end{table*}

\acknowledgments % equivalent to \section*{ACKNOWLEDGMENTS}       
This research is supported by the Director, Office of Science, Office of High Energy Physics of the U.S. Department of Energy under Contract No. DEAC0205CH1123, and by the National Energy Research Scientific Computing Center, a DOE Office of Science User Facility under the same contract; additional support for DESI is provided by the U.S. National Science Foundation, Division of Astronomical Sciences under Contract No. AST-0950945 to the National Optical Astronomy Observatory; the Science and Technologies Facilities Council of the United Kingdom; the Gordon and Betty Moore Foundation; the Heising-Simons Foundation; the National Council of Science and Technology of Mexico, and by the DESI Member Institutions. The authors are honored to be permitted to conduct astronomical research on Iolkam Du'ag (Kitt Peak), a mountain with particular significance to the Tohono O'odham Nation.
% References
\bibliography{report.bib} % bibliography data in report.bib
\bibliographystyle{spiebib} % makes bibtex use spiebib.bst

\section{References}
[1] Gunn, J.~E., Carr, M., Rockosi, C., et al. ``The Sloan Digital Sky Survey Photometric Camera ''AJ, 116, 3040 (1998)

[2] York, D.~G., Adelman, J., Anderson, J.~E., Jr., et al. ``The Sloan Digital Sky Survey: Technical Summary'' AJ, 120, 1579 (2000)

[3] Gunn, J.~E., \& Knapp, G.~R.\ ,``The Sloan Digital Sky Survey '' , ASPC, 43, 267 (1993)

[4] Aguilar, J., Leitner, D. et al., ``Dark Energy Spectroscopic Instrument (DESI) Fiber Positioner Production,'' in [Ground-based and Airborne Instrumentation for Astronomy VII], Proc. SPIE 10706-228 (2018). 

[5] Silber, J. et al., ``Dark Energy Spectroscopic Instrument (DESI) Focal Plane System,'' in [Ground-based and Airborne Instrumentation for Astronomy VII], Proc. SPIE 10702-311 (2018. )

[6] Schubnell, M., et al., ``DESI Fiber Positioner Testing and Performance, ''  in [Ground-based and Airborne Instrumentation for Astronomy VII], Proc. SPIE 10706-79 (2018).

%[6] Honscheid, K. et al., ``Status and Early Testing of the DESI Readout and Instrument Control System,'' in [Ground-based and Airborne Instrumentation for Astronomy VII], Proc. SPIE 10706 (2018). 

\end{document}